\documentclass[a4paper,11pt,usenames,svgnames]{article}
\usepackage{pos}
\usepackage{colortbl}
\DeclareRobustCommand{\Eq}[1]{Eq.~\eqref{eq:#1}}
\DeclareRobustCommand{\eq}[1]{eq.~\eqref{eq:#1}}

\DeclareRobustCommand{\fig}[1]{Fig.~\ref{fig:#1}}

\DeclareRobustCommand{\sec}[1]{Sec.~\ref{sec:#1}}

\DeclareRobustCommand{\tbl}[1]{Table~\ref{tbl:#1}}

\newcommand{\cC}{\ensuremath{\mathcal{C}}}
\newcommand{\cH}{\ensuremath{\mathcal{H}}}
\newcommand{\cPT}{\ensuremath{\mathcal{PT}}}
\newcommand{\cCK}{\ensuremath{\mathcal{CK}}}
\newcommand{\cP}{\ensuremath{\mathcal{P}}}
\newcommand{\cT}{\ensuremath{\mathcal{T}}}
\newcommand{\cK}{\ensuremath{\mathcal{K}}}
\newcommand{\cM}{\ensuremath{\mathcal{M}}}

\title{Finite-density QCD, $\cPT$ symmetry, and exotic phases}

\author[a]{Moses A.  Schindler}
\author*[b]{Stella T. Schindler}
\author[a]{Michael C. Ogilvie}

\affiliation[a]{Department of Physics, Washington University\\
	1 Brookings Drive, St. Louis, MO 63130}

\affiliation[b]{Center for Theoretical Physics, Massachusetts Institute of Technology,\\
  77 Massachusetts Ave., Cambridge, MA 02139}

\emailAdd{stellas@mit.edu}

\abstract{We study the phase structure of effective models of finite-density QCD using analytic and lattice simulation techniques developed for the study of non-Hermitian and $\cPT$-symmetric QFTs. Finite-density QCD is symmetric under the combined operation of the charge and complex conjugation operators $\cCK$, which falls into the class of so-called generalized $\cPT$ symmetries. We show that $\cPT$-symmetric quantum field theories can support patterned ground-state field configurations in the vicinity of a critical endpoint. We apply our methods to a lattice heavy quark model at nonzero chemical potential that displays patterning behavior for a range of parameters. We derive a simple approximate criterion for the formation of these patterns, which can be used with lattice results.
}

\FullConference{%
 The 38th International Symposium on Lattice Field Theory, LATTICE2021
  26th-30th July, 2021
  Zoom/Gather@Massachusetts Institute of Technology
}

\begin{document}
\maketitle

\section{Motivation}\label{sec:motivation}

Finite-density QCD belongs to the class of so-called non-Hermitian $\cPT$-symmetric systems due to its invariance under the $\cPT$-type operator $\cCK$, the simultaneous action of charge and complex conjugation \cite{Meisinger:2012va, Schindler:2021otf}. Since its inception in 1998 \cite{Bender:1998ke}, the field of $\cPT$ symmetry has had a wide impact on physics, generating thousands of experiments and applications across such diverse areas as optics, condensed matter, and quantum computing \cite{Miri2019, feng2017, Christodoulides2018, Ozdemir2019, doi:10.1142/q0178}. We bring knowledge of $\cPT$ symmetry to bear on lattice gauge theory, inspiring novel analytic and simulation techniques for finite-density models \cite{Ogilvie:2018fov, Medina:2019fnx, Schindler:2021otf}. Our findings indicate that finite-density QCD may have complicated phase behavior associated with the Polyakov loop, including inhomogeneous phases and sinusoidally-modulated Polyakov loop correlators near the critical endpoint. 

It has long been known that inhomogeneous behavior might be present in order parameters associated with quarks  \cite{Fukushima:2010bq}, including LOFF phases in color superconductors \cite{Alford:2007xm} and chiral spirals \cite{Kojo:2009ha}. 
It was first suggested in 2011 that propagators might exhibit sinusoidal modulation based on a flux tube model \cite{Patel:2011dp, Patel:2012vn}, which has a real action but is equivalent to a $Z(3)$ spin model with complex magnetic fields.
A number of phenomenological continuum effective field theories of finite-density QCD also exhibit a region of parameter space characterized by sinusoidally-modulated Polyakov loop correlators \cite{Nishimura:2014rxa, Nishimura:2014kla}. 
In Polyakov-Nambu-Jona Lasinio models, such modulation is known to occur near the critical endpoint in the $\mu-T$ plane \cite{Nishimura:2016yue}.
Strong-coupling lattice expansions of Polyakov loop correlators in the presence
of static quarks at finite density display sinusoidal modulation \cite{Nishimura:2015lit},
and there is strong evidence for this behavior in a $Z(3)$ model through simulation and mean field theory \cite{Akerlund:2016myr}.

In these proceedings, we use methods motivated by $\cPT$ symmetry to help explain the appearance of these exotic phenomena.
We develop a straightforward analytic probe for finding the phase structure of scalar QFTs, which we confirm numerically for a two-component theory using a new algorithm to circumvent the sign problem.
We then apply this analytic technique to a simple effective model of finite-density QCD.

\section{Brief introduction to $\cPT$ symmetry} 
			
A system is said to have a $\cPT$-type symmetry if it is invariant under the combined operations of any given linear operator ("$\cP$") and antilinear operator ("$\cT$").\footnote{This notation need not refer specifically to the parity inversion or time reversal operators of particle theory.} A prototypical $\cPT$-symmetric system is a Schr{\"o}dinger equation $\cH = -\partial_x^2+V(x)$ with a complex potential like $x^2(ix)^\epsilon$ satisfying
\begin{equation}\label{eq:pt-potential}
	V(x) = V^*(-x).
\end{equation}
$\cH$ is not Hermitian, but it is invariant under the simultaneous action of $\cP: x\to-x$ and $\cT: i\to-i$. An imaginary term in a potential signifies energy exchange between a system and its surroundings. Thus, $\cPT$ symmetry indicates balanced energy flow into and out of the system \cite{Bender:1998ke}. 

Each eigenvalue of a $\cPT$-symmetric operator $\cH$ must be either real or part of a complex-conjugate pair. When the eigenvalue spectrum is entirely real, we say that $\cH$ has an {\it unbroken} $\cPT$ symmetry.\footnote{This terminology has no direct connection to spontaneous symmetry breaking.} Unbroken $\cPT$ Hamiltonians exhibit analogues of traditional Hermitian properties such as unitary time evolution, orthogonality \cite{Bender:2007nj}, completeness \cite{Weigert:2003py}, and interlacing eigenfunctions \cite{Bender:2000wj, Schindler:2017suy}. These properties break down in a characteristic manner when symmetry is broken and one or more eigenvalue pairs becomes complex. 

The symmetry in \eq{pt-potential} is reminiscent of the fermion determinant of QCD at real nonzero $\mu$, which respects
\begin{equation}\label{eq:fermion-det}
	\det \cM (\mu) = \det \cM^*(-\mu).
\end{equation}
\Eq{fermion-det} is the source of the lattice QCD sign problem. It is also manifestly symmetric under the simultaneous action of the linear charge operator $\cC$ and antilinear complex conjugation $\cK$, a symmetry of the $\cPT$ type \cite{Schindler:2021otf}. Finite-density QCD is thus amenable to study using $\cPT$ methods.

\section{Phase structure of complex QFTs}\label{sec:phase-analytics}
We can determine the phase structure of a complex QFT by doing a conventional stability analysis of the propagator \cite{Medina:2019fnx, Schindler:2021otf}. Let us consider a generic theory with multiple scalar fields
\begin{equation}\label{eq:generic-action}
S[\phi_a(x),\, \chi_b(x)] = \frac{1}{2}(\partial_\mu \phi_a)^2 + \frac{1}{2}(\partial_\mu\chi_b)^2 + V(\phi_a,\chi_b),
\end{equation}
where we allow the potential $V(\phi_a,\chi_b)$ to have complex couplings and masses. We can write the momentum-space propagator as $G(p^2) = \frac{1}{p^2 + \cM}$, with mass matrix
\begin{equation}\label{eq:mass-matrix}
\cM = \left.\left(\begin{array}{cc} \frac{\partial^2 V}{\partial\phi_a^2} & \frac{\partial^2 V}{\partial\phi_a\partial\chi_b} \vspace{1 mm}\\ \frac{\partial^2 V}{\partial\chi_b\partial\phi_a} & \frac{\partial^2 V}{\partial\chi_b^2}\end{array}\right)\right\vert_{\phi_{a}^0,\,\chi_{b}^0}
\end{equation}
evaluated at homogeneous minima $\phi_a(x) = \phi_{a}^0$ and $\chi_b(x) = \chi_{b}^0$. 

In a conventional QFT, $\cM$ has nonnegative eigenvalues. In position space, the propagator decays exponentially, and thus the theory supports a stable homogeneous ground state. On the other hand, a generic Lagrangian with arbitrary complex terms has a non-Hermitian $\cM$ with a mix of positive, negative, and complex eigenvalues. In general, the position-space propagator would grow exponentially, signaling complete instability of the theory. 

Non-Hermitian Lagrangians with $\cPT$-type symmetries are better behaved.
Let us consider \eq{generic-action} with a $\cPT$-symmetric action satisfying 
\begin{equation}\label{eq:pt-action}
S[\phi_a(x),\chi_b(x)] = S^*[\phi_a(x),-\chi_b(x)].
\end{equation}
The potential $V$ and mass matrix $\cM$ obey the same symmetry as \eq{pt-action}.  
Thus, eigenvalues of $\cM$ are real or complex-conjugate pairs. 
Just like in a Hermitian system, when all eigenvalues of $\cM$ are real and positive, the position-space propagator decays exponentially, and we observe a normal homogeneous ground state. 
Likewise, when an odd number of eigenvalues are negative, the propagator blows up exponentially, and the QFT is unstable.

\setlength{\tabcolsep}{0.6em}
{\renewcommand{\arraystretch}{1.3}
	\begin{table}
		\begin{center}
			\begin{tabular}{c|c|c}
				{\bf Eigenvalues $E_i$ of $\cM$} & {\bf Position-space propagator behavior} & {\bf Region}\\
				\hline
				\rowcolor{AntiqueWhite}All positive & Exponential decay & Normal\\\hline
				\rowcolor{MistyRose}Odd number of $E_i < 0$ & Exponential growth & Unstable\\\hline
				\rowcolor{LightCyan}Some $E_i = E_j^*$ & Sinusoidally-modulated exponential & $\cPT$ broken\\\hline
				\rowcolor{Honeydew}Even number of $E_i<0$& Homogeneous solution unstable at some $p \neq 0$ & Patterned vacuum\\
			\end{tabular}
		\end{center}
		\caption{Stability analysis of the $\cPT$-symmetric action \eq{generic-action} with the condition \eq{pt-potential} by examination of its mass matrix \eq{mass-matrix}. Note the two behaviors not observed in a conventional QFT: the $\cPT$-broken region, which has a sinusoidally-modulated propagator and a stable homogeneous ground state, and the patterned region, which is characterized by a Lifshitz instability and a stable inhomogeneous ground state.}\label{tbl:table}
	\end{table}
}

$\cPT$-symmetric QFTs exhibit two behaviors that do not occur in conventional QFTs, which we can observe directly from $\cM$. 
First, when $\cM$ has one or more complex-conjugate eigenvalue pairs, the propagator decays exponentially in position space, but with modulation from a sine term. 
Nonetheless, this region still maintains a stable homogeneous ground state. 
Second, $\cM$ can have an even number of negative eigenvalues. In this case, the homogeneous solution is unstable to fluctuations for some momenta $p \neq 0$, but it is stable for $p = 0$. Instability of a homogeneous phase need not indicate instability of the full theory; rather, here it merely signifies that the stable phase is inhomogeneous: i.e., a so-called Lifshitz instability. We summarize these behaviors in \tbl{table}. The case $E=0$ gives massless modes; in some models, $\cPT$ Goldstones are obtained \cite{Fring:2021zci,Mannheim:2021kjs}.

\subsection{Example: $\cPT$-symmetric $\phi^4$ model}
Let us develop intuition for $\cPT$-QFTs by applying the mass-matrix analysis to a simple theory, and then cross-checking our results with simulations. Consider a two-component $\cPT$-symmetric extension of the conventional $\phi^4$ model \cite{Medina:2019fnx}: 
\begin{equation}\label{eq:phi4}
	S[\phi(x), \chi(x)] = \frac{1}{2}(\partial_\mu\phi)^2 + \frac{1}{2}(\partial_\mu\chi)^2 + \frac{1}{2}m_\chi^2\chi^2 -ig\phi\chi + \lambda(\phi^2-v^2)^2 + h\phi.
\end{equation}
Here, we have a massless field $\phi(x)$ sitting in a double-well potential, coupled by an imaginary strength $ig$ to a massive $\cPT$-symmetric field $\chi(x)$. We can adjust the heights of the two wells and break their symmetry with the source term $h\phi$. Plugging \eq{phi4} into \eq{mass-matrix}, we can calculate the mass matrix of $S$:
\begin{equation}\label{eq:phi4-mass-matrix}
\cM = \left(\begin{array}{cc} 4\lambda(3\phi_0^2-v^2) & ig \vspace{1 mm}\\ ig& m_\chi^2\end{array}\right),
\end{equation}
where we take $\phi(x) = \phi_0$ as the tree level minimum of $V$. The eigenvalues of $\cM$ are
\begin{equation}
E_{\pm} = 2\lambda(3\phi_0^2-v^2) + \frac{1}{2}m_\chi^2 \pm \frac{1}{2}\sqrt{[4\lambda(3\phi_0^2-v^2)-m_\chi^2]^2-4g^2},
\end{equation}
which can become complex for sufficiently large couplings $g$. Note that a direct mass matrix analysis requires us to work with $\langle \phi \rangle = \phi_0$ as a parameter, rather than just the more natural coupling constants $\lambda,\,g,\,h,\,v^2$ in $S$. 

\begin{figure}
\begin{center}
\includegraphics[width = 2.5in]{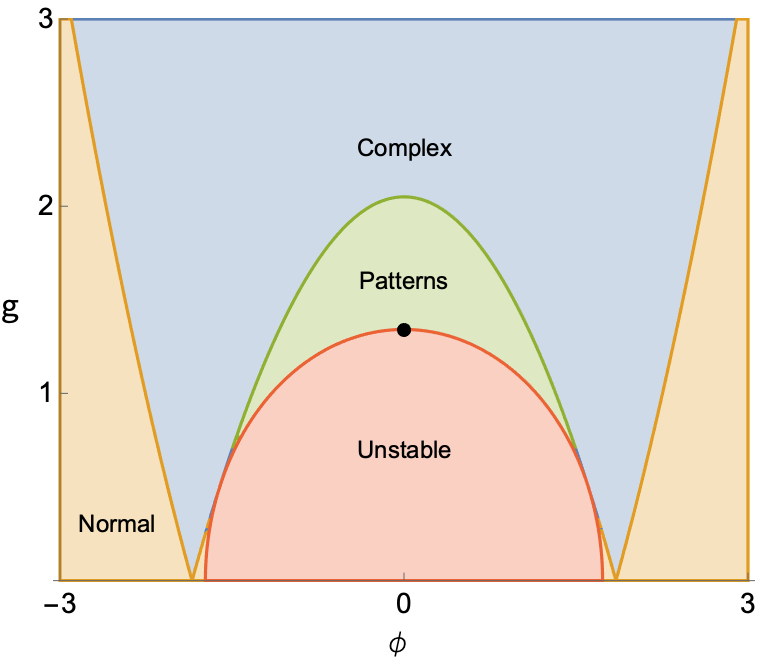}\qquad
\includegraphics[width = 2.5in]{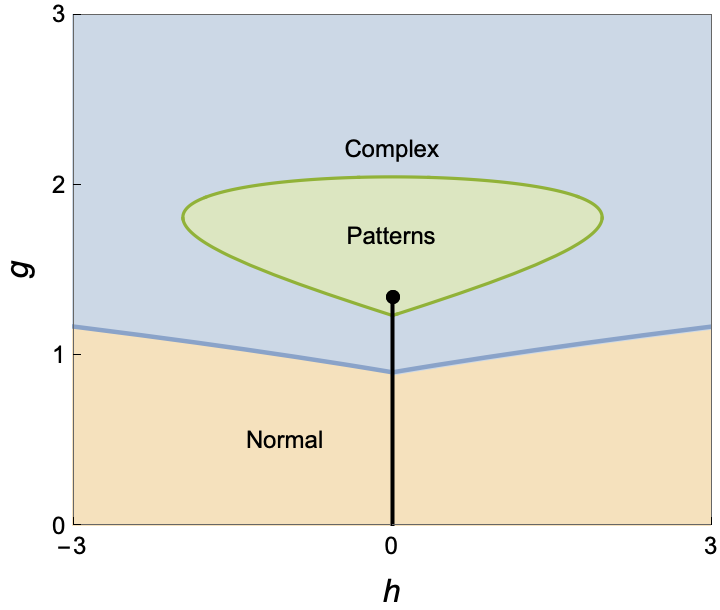}
\end{center}
\caption{Phase diagram of \eq{phi4} determined analytically at tree level, and plotted in (a) the $\langle \phi \rangle$-$g$ plane and (b) the $h$-$g$ plane. We fix $m_\chi^2 = 0.5$, $\lambda = 0.1$, and $v = 3$.} \label{fig:phi4-phases}
\end{figure}

We begin by discussing a familiar case: decoupling the fields with $g=0$, leaving us with a conventional $\phi^4$ theory. From \eq{phi4-mass-matrix}, we see that $\phi(x)$ has mass $E = 4\lambda(3\phi_0^2-v^2)$. Thus, in the $\langle \phi \rangle-v^2$ plane, we observe an unstable phase in the region $3\langle \phi\rangle^2<v^2$, metastable states centered around $\langle \phi \rangle = 0$, and a second-order critical point at $(\langle \phi \rangle,\, v^2) = (0,\,0)$. Taking a Legendre transform gives us the phase diagram of \eq{phi4} in the $h-v^2$ plane: the metastable and unstable states of the $\langle\phi\rangle - g$ plane are replaced by a cut along the $h=0$ axis for all positive values of $v^2$. 

Now, we examine nonzero $g$ values, varying $g$ and $h$ while holding $\lambda$ and $v^2$ fixed. We again start by analyzing $\cM$ in the $\langle \phi \rangle-g$ plane. In \fig{phi4-phases}, as expected we see four phases: stable, unstable, complex, and patterned. A spinodal line separates the unstable and patterned regions. A second-order critical point lies on the spinodal line at $\langle\phi\rangle=0$. 

We can Legendre-transform these results into the $h-g$ plane. As in the normal $\phi^4$ model, metastable states in the complex (blue) phase and unstable states (red) disappear and collapse into a critical line running up the $h=0$ axis. This line is a first-order phase transition terminating in a second-order critical endpoint; i.e., $\langle \phi \rangle$ exhibits a discontinuity as we vary $h$ past $h=0$. Note that after the Legendre transform, the critical endpoint has relocated from the edge of the patterned region to its interior. 

\subsection{Simulations}
\begin{figure}
	\begin{center}
		\includegraphics[width = 4.5 in]{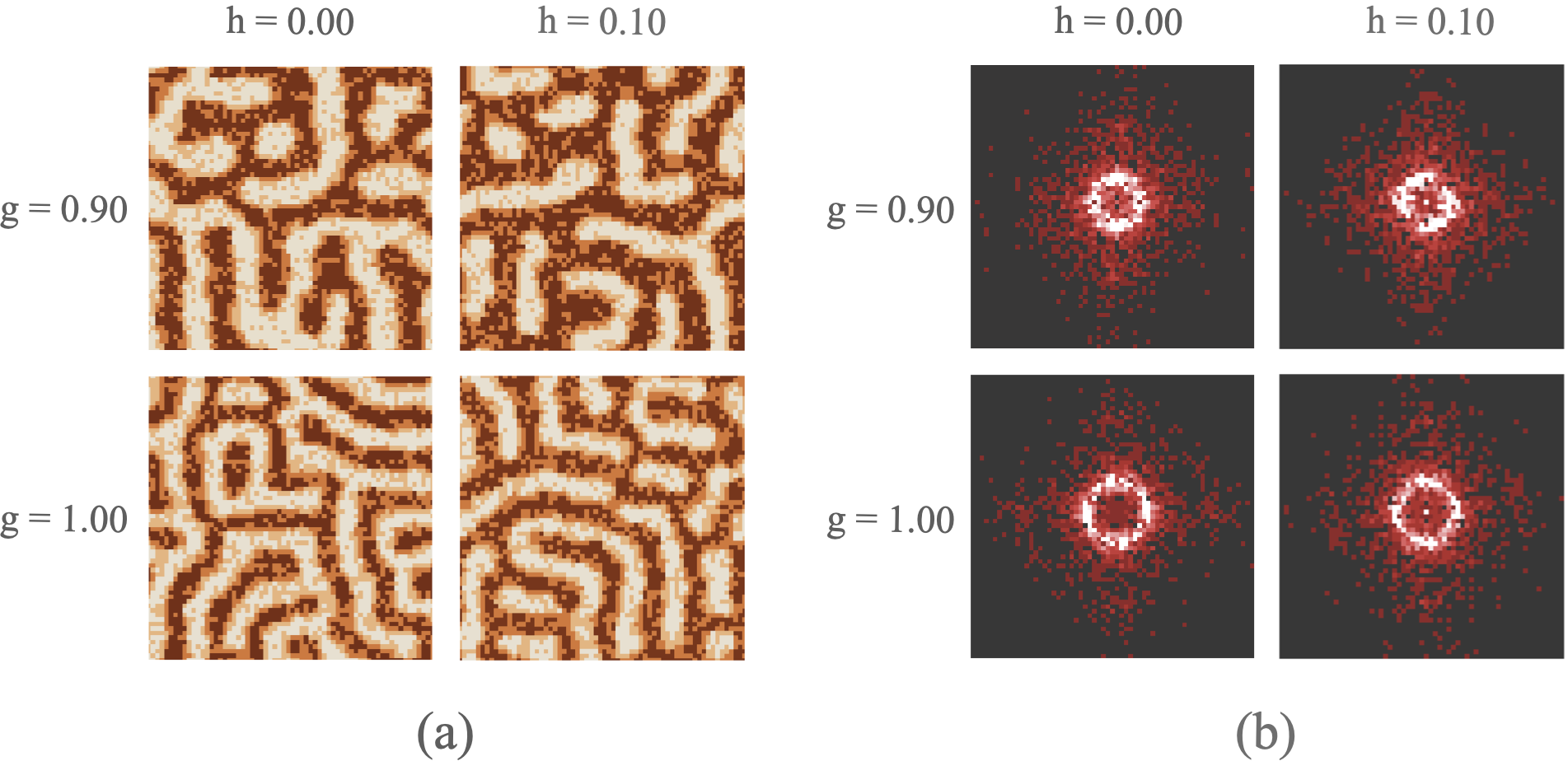}
	\end{center}
	\caption{Snapshots of \eq{simulations} simulation results. Pictured: (a) configurations $\phi(x)$ and (b) their position-space Fourier transform $\tilde{\phi}(k)$. We plot a range of values $(g,h)$ at fixed $m_\chi^2 = 0.5$, $\lambda = 0.1$, and $v = 3$. Each configuration was obtained on a $64^2$ lattice after a hot start followed by 20,000 sweeps. The color scale in (a) runs from -3 to 3, from dark to light. The color scale in (b) runs from 0 to 10, from dark to light; any lattice point with magnitude greater than to is set to 10 for ease of visualization. Note that patterned configurations evolve smoothly as we adjust $(g,h)$, and always correspond to ring-shaped configurations in Fourier space.}\label{fig:configs}
\end{figure}

Medina and Ogilvie have developed a technique to recast $\cPT$-QFTs into a fully real form \cite{Ogilvie:2018fov}: all we need to do is take a Fourier transform of appropriate fields within the path integral. To simulate $S$ in \eq{phi4}, we make such a transform with respect to the field $\chi$ ({\it not} the position-space variable $x$). The dual action $\tilde{S}$ is real, positive, and thus simulatable:
\begin{equation}\label{eq:simulations}
	\tilde{S}[\phi(x), \pi(x)] = \frac{1}{2} (\partial_\mu\phi)^2 + \frac{1}{2} \pi_\mu^2 + \frac{1}{2m_\chi^2}\left[\partial\cdot\pi - g\phi \right]^2 + \lambda(\phi^2-v^2)^2 + h\phi.
\end{equation}
In \fig{configs}, we display a small sample of lattice simulation results. We notice that the region of pattern formation in lattice simulations using $h$ as a parameter agrees well with the region we determined using tree-level analytics using $\langle \phi \rangle$ as a parameter in \fig{phi4-phases}. 

The nature of the patterned field configurations strongly suggests that the patterned region comprises a single thermodynamic phase. As we vary the parameters $(g,h)$, the pattern morphologies evolve smoothly. When we take a Fourier transform of a configuration from $\phi(x)$ to $\tilde{\phi}(k)$, we see that patterned configurations in $x$-space become ring-shaped in $k$-space. The only evidence for a sharp thermodynamic phase transition in the patterned region is at $h=0$ for low values of $g$, which is where we have analytically determined the critical line and critical endpoint to be located. Along this line, $\langle \phi \rangle$ jumps in a first-order transition that terminates at a second-order point.

\section{Phase structure of a QCD-like model}
Let us now apply these methods to a simple effective model of finite-density QCD, a heavy quark theory \cite{Schindler:2021otf}. Consider the action
\begin{equation}\label{eq:heavy-quark}
S = -2N_CN_F\sum_{\vec{x}} \log \left[1 + ze^{\chi(\vec{x}) + i\phi(\vec{x})} \right] + \frac{1}{2\beta}\sum_{\vec{x},\,\vec{y}}\left[\chi(\vec{x})V_A^{-1}(\vec{x},\vec{y})\chi(\vec{y}) + \phi(\vec{x})V_R^{-1}(\vec{x},\vec{y})\phi(\vec{y})\right].
\end{equation}
The first term is a mean-field heavy quark determinant; we parametrize the Polyakov loop as $P = e^{\chi + i\phi}$ and absorb parameters into a coefficient $z = e^{\beta(\mu-m)}$. The second term characterizes Polyakov loop interactions with a potential split into attractive and repulsive pieces $V = V_A + V_R$. 

\begin{figure}
	\begin{center}
		\includegraphics[width = 2.5 in]{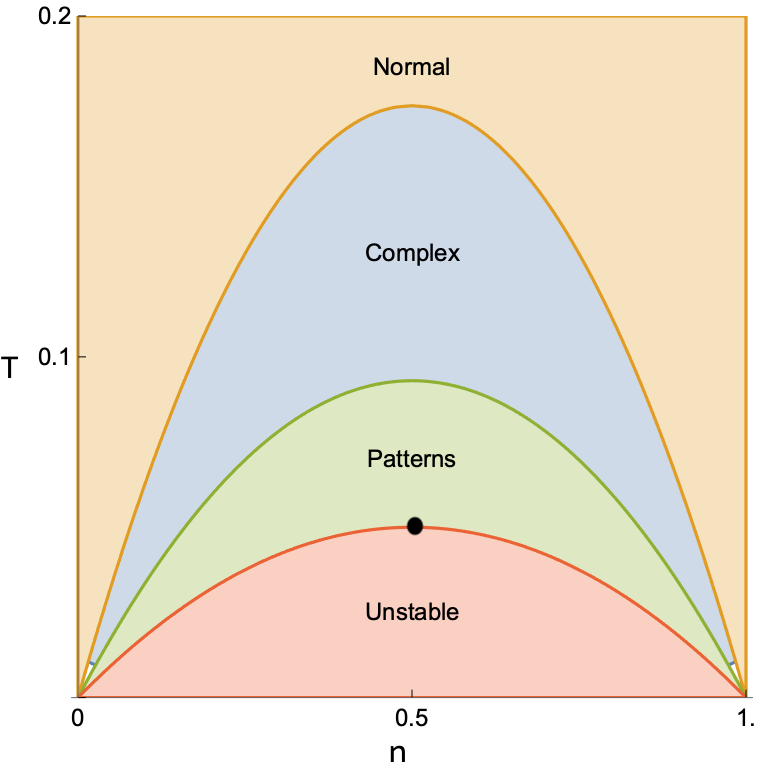}\qquad
		\includegraphics[width = 2.5 in]{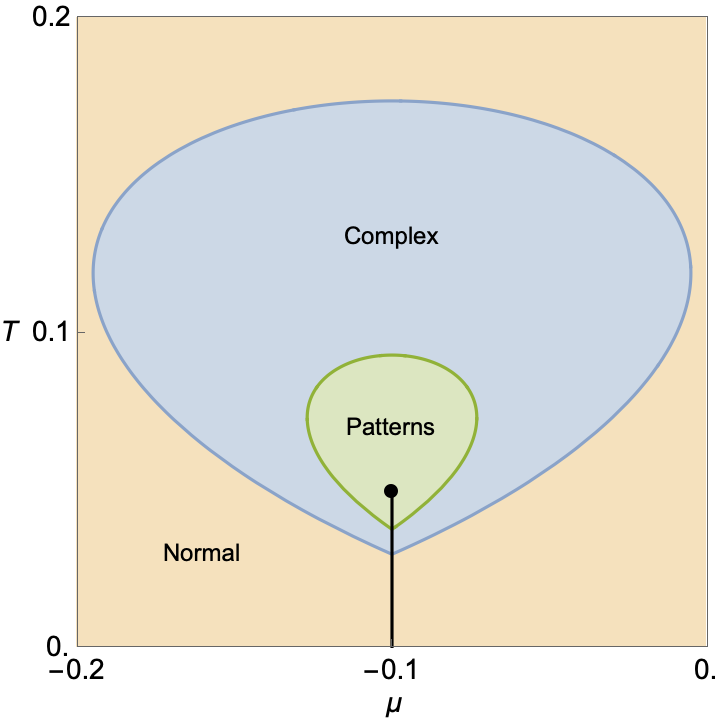}
	\end{center}
	\caption{Plot of the analytically-determined heavy quark model phase diagram in (a) the $n$-$T$ plane and (b) the $\mu$-$T$ plane. The patterned region consists of bubbles of deconfined matter floating in a sea of confined matter, and vice versa. Note that in (b) the critical endpoint is surrounded by a patterned phase, then the sinusoidally-modulated region, which is bounded by a disorder line. These behaviors may be physically observable.}\label{fig:heavy-quark}
\end{figure}

We can determine the phase diagram of \eq{heavy-quark} straightforwardly using the procedure outlined in \sec{phase-analytics}. It is instructive to first examine a simple choice of potentials: Yukawa interactions $V_i^{-1} = \frac{1}{g_i^2}(\nabla^2 + m_i^2)$ with $i = \phi, \chi$. We plot the phase diagram of the Yukawa model in both the fermion density $n$ vs. $T = 1/\beta$ plane, as well as in the chemical potential $\mu$ vs. $T$ plane in \fig{heavy-quark}.  

The phase diagram structure looks remarkably similar to that of \eq{phi4}: around the critical endpoint, we find a patterned region, surrounded by a sinusoidally-modulated region, which is in turn bounded by a disorder line. These behaviors are indeed universal in $Z(2)$ pattern formation, which raises the possibility that we might see similar phenomena around the critical endpoint of finite-density QCD. The inhomogeneous phase would consist of patterns of deconfined phase within a sea of confined phase, and vice versa. 

We can calculate the phase diagram for the full model as in \tbl{table}, and  extend the results to the continuum. We write the criterion for stability against pattern formation as
\begin{equation}
	1 + \beta \mathcal{X}_Q \tilde{V}_{qq}(k) > 0,
\end{equation}
where $\mathcal{X}_Q$ is the quark number susceptibility and $\tilde{V}_{qq}$ is the quark-quark potential. Note that all terms in this stability criterion are in principle computable on the lattice.

\section{Outlook}
We believe that the $\cPT$-symmetric framework holds great promise in understanding critical behavior in field theories with sign problems, including finite-density QCD. 
There are two natural directions in which to proceed. 
First, we can directly use the techniques shown in these proceedings to tackle more complicated models that incorporate more features of finite-density QCD. 
For example, dropping the mean-field approximation implicit in \eq{heavy-quark} allows for a more precise determination of the critical point of a heavy quark model. 
Also, we can examine the effect of adding in canonical features of QCD like triality and chirality, and see their effect on phase structure.  
Second, it is important to extend our suite of $\cPT$-inspired analytic and algorithmic techniques to accommodate a wider span of field theories. 
We are pursuing both lines of research.

\bibliographystyle{JHEP}
\bibliography{stella-edits.bib}

\end{document}